# Symmetry-driven phonon confinement in 2D halide perovskites

Mustafa Mahmoud Aboulsaad[a], Olivier Donzel-Gargand[b], Rafael B. Araujo[a,*], Tomas Edvinsson[a,*]

[a] Department of Materials Science and Engineering, Solid State Physics, Uppsala University, Box 35, 75103 Uppsala, Sweden

[b] Department of Materials Science and Engineering, Solar Cell Technology, Uppsala University, Box 35, 75103 Uppsala, Sweden

[*] Corresponding authors: rafael.araujo@angstrom.uu.se , tomas.edvinsson@angstrom.uu.se

Abstract

Quantum confinement not only reshapes electronic states but also reorganizes the vibrational landscape of low-dimensional semiconductors. In halide perovskites, however, the role of confinement in governing symmetry effects on vibrational modes has remained unresolved. Here we synthesize 2D $CsPbBr_3$ nanoplatelets with atomically defined thicknesses for 2–5 monolayers (MLs) and perform exciton absorption and emission analysis, crystalline phase determination, and phonon analysis. The lowest dimensional structure (2 MLs) reveal a co-existence of cubic and orthorhombic structure, energetically converging to orthorhombic for 3 MLs and beyond. Through polarization-resolved Raman spectroscopy and first-principles theory for 2-5 MLs, a striking symmetry contrast is found: $B_{1g}$ modes intensify and evolve in line with the phonon-confinement model, while $A_g$ modes deviate, reflecting their distinct spatial localization. First principles calculations show that $B_{1g}$ vibrational modes largely reside in the xy-plane, Pb-Br-Pb units connect octahedra along the xy-direction with increased lattice dynamics as inner layers accumulate, whereas $A_{1g}$ vibrations couple to out-of-plane distortions and remain susceptible to surface disorder and finite-size effects. This symmetry-driven dichotomy provides a general framework for understanding phonon localization in layered halide perovskites. Beyond mechanism, we establish Raman fingerprints, particularly the $A_{1g}/B_{1g}$ intensity ratio in cross-polarized geometry, as a calibrated, non-destructive metrology for 2D nanoplatelet thickness through 2-5 MLs. These results bridge electronic and phonon confinement and highlight symmetry engineering as a route to understand and control phonons, excitons, and their interactions in low-dimensional optoelectronic materials.

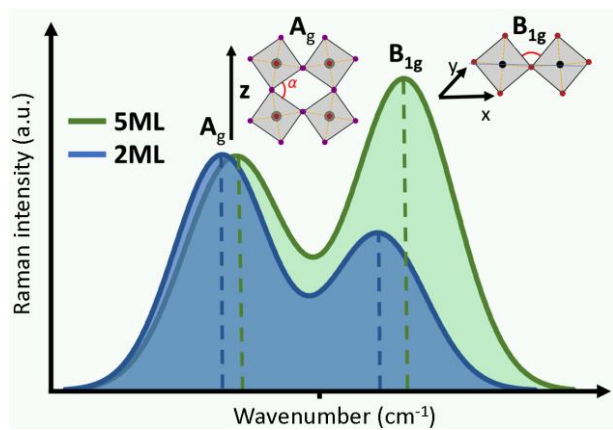

**Main**

Two-dimensional (2D) halide perovskite nanoplatelets (NPLs) have emerged as transformative materials in optoelectronics, promising applications in light-emitting diodes (LEDs), lasers, and photodetectors. Their appeal lies in highly tunable electronic and optical properties engineered for specific device requirements. Early tuning approaches emphasized compositional engineering of the perovskite lattice, such as halide substitution or variation of the primary cation.[1] More recently, attention has shifted to exploiting quantum confinement from reduced thickness. Confinement alters the electronic structure by spatially localizing charge carriers, reshaping the density of states (DOS), exciton binding energy, and bandgap.[2] A key manifestation is the systematic blue shift in emission with decreasing thickness, reflecting reduced dimensionality.[3–5] Thus, NPLs provide a versatile platform for both device applications and fundamental studies of size-dependent quantum phenomena.

Extensive research has examined confinement effects on electronic structure and photoluminescence (PL). Reports consistently show that reduced thickness enhances PL and photoluminescence quantum yield (PLQY).[6–8] Density functional theory (DFT) confirms that thinning increases exciton binding energy via spatial confinement and dielectric modifications.[9] Yet, vibrational properties across different thicknesses remain less understood, with only limited studies available.[10–14] Existing work often focuses on electron–phonon or exciton–phonon coupling, leaving open how confinement directly modulates vibrational transversal optical (TO) modes and their spectroscopic signatures. Filling this gap is critical, as TO modes are the dominant number of modes, where vibrational dynamics affect structural and thermal properties, and provide diagnostic tools for confinement.

Raman spectroscopy offers a powerful analysis tool. As a non-destructive probe of phonon modes, interlayer interactions, and lattice symmetry, Raman has been indispensable in 2D materials such as graphene, where the G (~1580 cm⁻¹) and the overtone of the first-order forbidden D mode (~1350 cm⁻¹) appearing as the 2D (~2700 cm⁻¹) band, enabling the ability to distinguish mono- to tri-layers from multilayers and graphite.[15] In nanoparticles, confinement and symmetry lowering relax the $q \approx 0$ selection rule, enabling off-center phonons to contribute. This generates thickness-dependent frequency shifts, line-shape asymmetries, and altered intensity ratios of symmetry-allowed channels, establishing Raman as a precise probe of confinement in NPLs.[16–18] Similar size-dependent Raman responses are well documented in oxides such as $TiO_2$ and ZnO.[16–19] Beyond confinement, Raman is highly sensitive to anisotropy and symmetry breaking,[20,21] enabling layer-number identification and detection of subtle perturbations. Polarization-resolved approaches strengthen this capability: by isolating symmetry channels (co- vs cross-polarized geometries), one can quantify interlayer coupling[22] and surface localization, and derive symmetry-resolved metrics, such as cross/co-polarized ratios and line shape asymmetry parameter, that directly report on confinement and provide calibrated thickness readouts. In the last decades, this have been achieved for crystalline materials such as graphene and TMDs, mainly having hexagonal crystal structure with a small basis (2 atoms in the basis). Here, either intensity ratio changes (graphene) or line-shape/frequency shifts (TMDs) are utilized, while the situation for materials with a larger basis and their precise polarization changes as a function of 2D confinement have been elusive.

Here, we synthesized $CsPbBr_3$ NPLs with defined thicknesses (2–5 monolayers) and comparison with nanocrystals (NCs) as bulk-like references. The system contains a complex basis with internal coordinates corresponding to (0,0,0) Pb, (½,0,0), (0, ½, 0), (0,0,½) for the halogens, and (½½,½) for the A-site cation Cs for a cubic system, with naturally skewed halogen coordinates and increased number of atoms in basis for the orthorhombic case. Structural characterization by X-ray diffraction and electron microscopy confirmed orthorhombic frameworks with discrete layer counts. Optical absorption and PL revealed the expected confinement-induced blue shift and systematic linewidth changes. Building on this foundation, we employed polarization-resolved Raman spectroscopy with DFT to resolve how vibrational symmetry evolves under confinement. We observed pronounced thickness-dependent enhancement of $B_{1g}$ modes relative to $A_g$ modes. $B_{1g}$ evolution follows phonon-confinement expectations, showing red shifts and asymmetric low-frequency tails under stronger confinement, whereas $A_g$ trends deviate. Eigenvector analyses clarify this: Pb–Br–Pb bridge octahedra along xy in $B_{1g}$ modes, strengthening as inner layers increase, while $A_g$ vibrations couple to out-of-plane (z) modes and are therefore highly sensitive to surface termination, finite thickness, and disorder.

Finally, exploiting Raman's sensitivity, we demonstrate that thickness-induced confinement in halide perovskite NPLs can be tracked with high precision. Vibrational-mode evolution provides a spectroscopic fingerprint linking structure to optical response. Polarization-resolved measurements disentangle $A_g$ and $B_{1g}$ channels, quantify interlayer coupling and surface localization, and yield robust observables such as the $B_{1g}/A_g$ intensity ratio and a lineshape asymmetry parameter. These symmetry-resolved metrics establish Raman spectroscopy as a calibrated, non-destructive metrology for identifying NPL thickness from a single spectrum. To further assess the generality of this framework, we extended our analysis to iodine-based perovskite nanoplatelets and observed the same symmetry-governed trends. We further validate our model to investigate the influence of post-synthetic treatments on the colloidal stability and properties of the nanoplatelets. This validation indicates that the model is broadly applicable across halide perovskite 2D systems and, more generally, to other layered materials with symmetrically related Raman modes. In summary, our work bridges electronic and vibrational confinement in halide perovskites and positions Raman spectroscopy, complemented by PL and first-principles theory, as a powerful framework for understanding and engineering phonon–carrier interactions in low-dimensional semiconductors.

**Structural analysis**

CsPbBr$_3$, like other perovskites, adopts multiple crystal phases across a wide temperature range. At room temperature, single-crystal and polycrystalline forms are orthorhombic; upon heating, single crystals convert to tetragonal at 126 °C, while polycrystals become cubic at 130 °C.[23,24] In NCs, phase behavior diverges from bulk. Dependent on the conditions, CsPbBr$_3$ NCs exhibit alternative phases at room temperature with transitions within specific windows.[25,26] For instance, NCs orthorhombic at room temperature may transform to cubic at 117 °C.[23] A tetragonal I4/mcm phase is also reported in single crystals[27] and nanocrystal thin films from 7 to 360 K.[28]

The crystal structure of nanosized CsPbBr$_3$ remains debated. Some XRD studies report cubic NCs, others orthorhombic. Cottingham and Brutchey[29] used synchrotron PDF analysis of 6.5 and 12.5 nm NCs, identifying orthorhombic symmetry. Brennan et al. examined ~10 nm NCs with TEM, finding mostly cubic with minor orthorhombic contributions.[30] Electron microscopy confirms heterogeneity at the single-particle level,[31,32] but the prevailing view is that orthorhombic is the stable phase under standard conditions.[23,29,33]

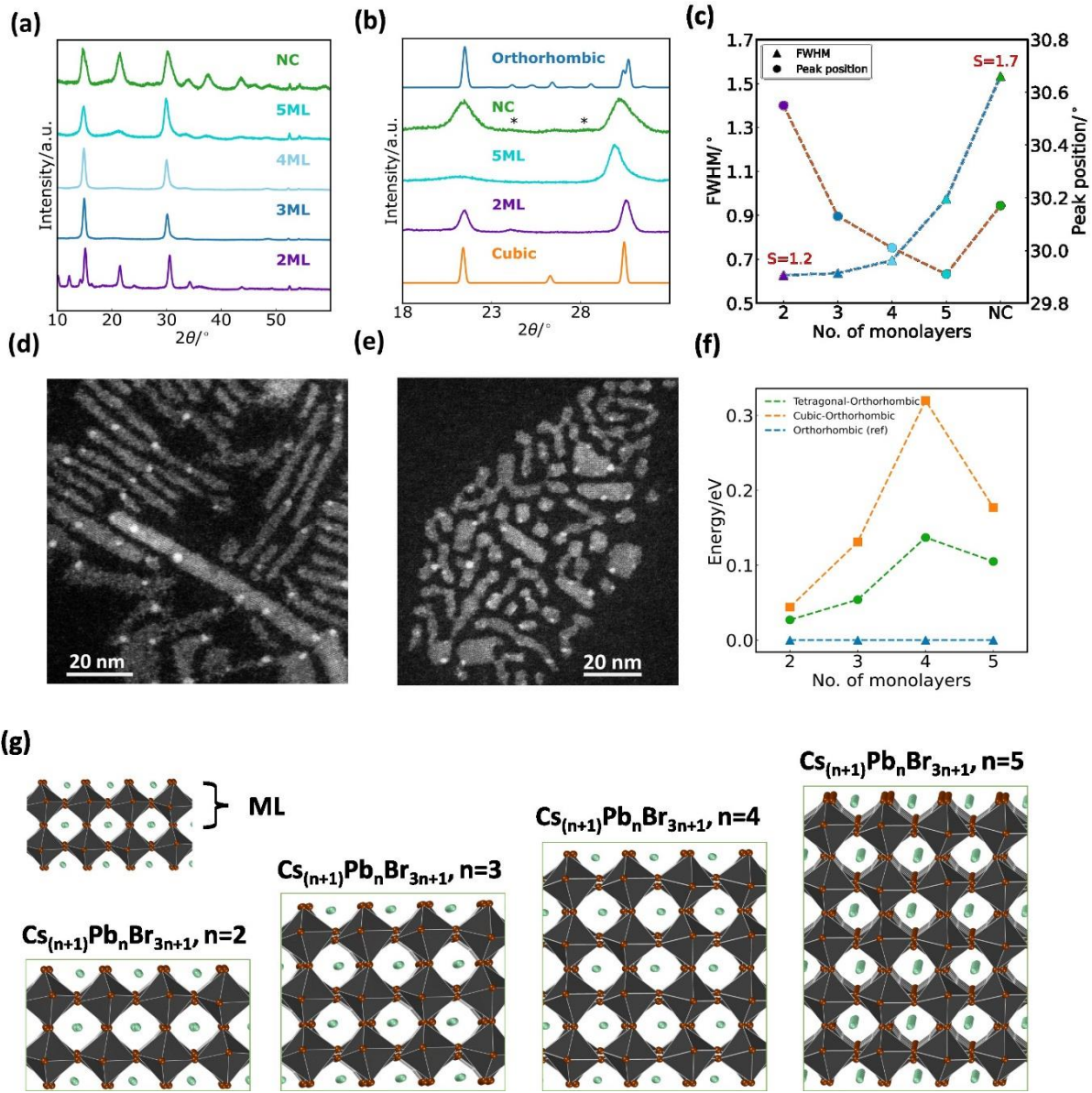

Figure 1: (a) GIXRD for the nanocrystal systems indicating the peak of interest at 31°. (b) orthorhombic features of NCs compared with cubic and orthorhombic phases of $CsPbBr_3$, as well as 2 and 5 MLs NPLs. (c) FWHM and peak shift analysis of the diffraction peak at 31°. S is the symmetrical parameter, where S=1 indicates the highest symmetry. The full method for the calculation of the symmetry factor is in the supplementary information. (d) and (e) STEM annular dark-field images showing the agglomerations of 3 and 5 MLs NPLs, respectively (f) Relative total energies of $CsPbBr_3$ NPls with 2–5 MLs, referenced to the orthorhombic phase. The cubic–orthorhombic (orange squares) and tetragonal–orthorhombic (blue circles). Here, the ground state case from the minima hopping was employed for all cases. (g) Orthorhombic structural models for NPLs calculations.

Our XRD measurements show primary peaks at 15° and 31° in 2θ (Figure 1a). Orthorhombic $CsPbBr_3$ should be identifiable by its denser diffraction pattern than cubic.[23] However, Scherrer broadening merges closely spaced reflections, producing broadened peaks. Still, orthorhombicity is evident through weak peaks (indicated with black stars in Figure 1b) absent in cubic.[23] Shoulders and asymmetric line shapes further support this assignment, especially in 5 ML and NC samples (Figure

1b,c).[32] Similar distortions occur in NPLs, previously confirmed to be orthorhombic with slight in-plane symmetry breaking.[34,35] Variations in intensity ratios are attributed to anisotropic orientation during drop-casting.[32] No $Cs_4PbBr_6$ impurity peaks were detected.[36]

The 2 ML sample deviates, showing additional low-angle peaks at 10.1°, 12.2°, 14.2°, 16.3°, and 18.4° in 2θ, corresponding to lamellar spacings of ~4.2 nm, characteristic of stacked NPLs, confirming preserved geometry.[37,38] Another trend is peak shifts to higher 2θ with decreasing thickness, attributed to increased surface energy.[39–42] Orthorhombic features, including a, b, c peaks, the 31° FWHM, and asymmetric shoulders, weaken with fewer monolayers. The symmetry factor S quantifies this asymmetry, which grows with thinning, indicating a gradual orthorhombic-to-cubic transition. NCs exhibit stronger asymmetry than 2 ML samples. A similar thickness-dependent evolution was reported by Wang et al., driven by Pb:Cs ratio increases.[43]

Minima hopping calculations of orthorhombic, cubic, and tetragonal phases (see SI) support these observations. Lowest-energy configurations were selected for each case. Energy differences between phases increase with layer number, stabilizing orthorhombicity (Figure 1g shows orthorhombic lowest-energy structures; Figure 1f relative energies), while tetragonal remains unfavorable. For 2 ML, the orthorhombic–cubic gap is 0.027 eV, close to $k_{BT}$ at room temperature (~0.026 eV), suggesting coexistence. At larger thickness, orthorhombicity dominates, confirming it as the more likely phase.

Figure 1d–e show TEM images for the formation of 3- and 5-MLs NPLs. Dark-field images reveal sparse, intensely bright nanodots that we assign to metallic Pb precipitates, consistent with electron-beam–induced halogen loss and radiolytic reduction of $Pb^{2+}$ to $Pb^0$ reported for lead-halide perovskites.[44] 4D-STEM diffraction from the same regions shows mixed cubic and orthorhombic features, in line with crystallographic heterogeneity at the single-particle level.[30] The material is quite sensitive to the electron beam, most of the lattice contrast disappearing after the first scan as shown in SI (Figure S2)[45]. Under continuous illumination, a few unidentified SAED rings weaken or vanish within seconds, giving evidence for beam-driven sensitivity; thus, TEM diffraction and high-resolution imaging should be interpreted with caution and acquired under low-dose conditions.[45] More details about beam sensitivity and the EM diffraction analysis can be found in the SI (Figures S2-8).

**Optical Analysis**

The PL spectra (Figure 2a) exhibit narrow peaks (FWHM: 12.7 nm for 2 ML to 21 nm for 5 ML) with emission shifting from 488 to 431 nm as the Cs/Pb ratio decreases. Under UV, this corresponds to a visible change from deep blue (2 ML) to cyan (5 ML), and to green emission at 512 nm (FWHM = 20 nm) for nanocubes, consistent with weak quantum confinement in 3D nanocubes.[5,46] Absorption spectra (Figure 2b) display a sharp onsets and Stokes shifts of 4–23 nm. At lower Cs/Pb ratios, pronounced excitonic peaks and step-like continuum absorption, characteristic of 2D semiconductors, are observed, reflecting increased exciton binding energy (EB) and reduced exciton dissociation. These excitonic features weaken with increasing thickness and vanish in nanocubes.

The PL blue shift at reduced Cs/Pb ratios originates from quantum confinement, which raises the continuum onset energy ($E_C = E_G + E_e + E_h$). Enhanced Coulomb interactions in confined dimensions increase EB, shifting EC upward. PL intensity rises exponentially from 2 MLs to 5 MLs (Figure 2c) due to more photoactive sites, while PLQY remains low; 4% (2 ML), 5% (3 ML), versus 27% for NCs (Figure 2d), likely due to trap states and large surface-to-volume ratios. These energies and PLQYs compare well with those reported by Bohn et al.[47], while the lower PLQYs in our synthesis indicate a slightly higher trap densities.

The PL trends agree with our dielectric-dependent excitonic analysis. Calculations for the orthorhombic phase reproduce the first excitonic peaks and match the experimental blue shift. Exciton binding energies decrease modestly with thickness (from 0.26 → 0.21 eV with n = 2 to 5), while band-structure shifts dominate. This agreement supports the orthorhombic phase, identified by XRD, as the structural ground state.

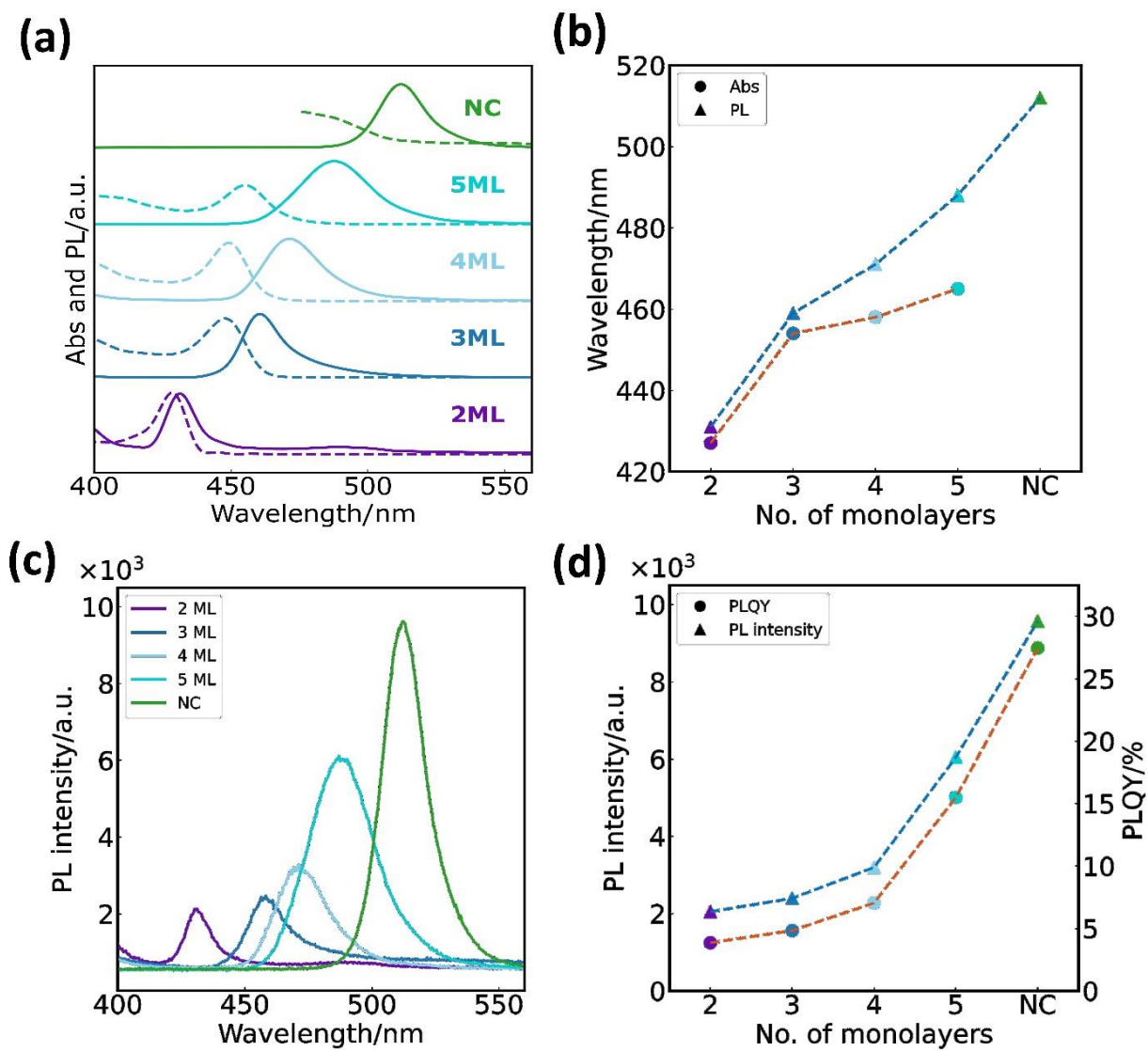

Figure 2: (a) Normalized UV-Vis and PL spectra, (b) Peak positions for the excitonic absorption/emission maxima, (c) Non-normalized PL spectra, and (d) PL and PLQY intensities.

**Vibrational analysis**

The Raman spectra of NPLs and NCs display fundamental bands centred at 47 cm$^{-1}$, 73 cm$^{-1}$, and 126 cm$^{-1}$ (Figure 3a) for all samples. The two lower-frequency bands dominate in intensity, while the 126 cm$^{-1}$ band is weaker. A clear enhancement of the 73 cm$^{-1}$ band relative to the others is observed when progressing from 2 ML to the NC case, quantified through the intensity ratio of the 47 cm$^{-1}$ and 73 cm$^{-1}$ bands (Figure 3b). In 2 ML NPLs, the 47 cm$^{-1}$ mode dominates, whereas in 3 ML the intensities are comparable. For thicker samples, the 73 cm$^{-1}$ mode remains stronger than the 47 cm$^{-1}$ mode. Additionally, a red shift of the 73 cm$^{-1}$ mode is observed from 2 ML to the NC case (Figure 3c), while the 47 cm$^{-1}$ band shows only a weak blue shift, evident between 5 ML and the NCs.

Vibrational modes were further analyzed using DFT. Calculations for the single-crystal orthorhombic phase (Pnma) showed no imaginary phonon frequencies, confirming stability (Figure 3d). The zone-center optical modes decompose as $\Gamma_{optic} = 7A_g + 8A_u + 5B_{1g} + 9B_{1u} + 7B_{2g} + 7B_{2u} + 5B_{3g} + 9B_{3u}$, of which $7A_g + 5B_{1g} + 7B_{2g} + 5B_{3g}$ are Raman active. Four dominant modes were predicted at 45, 65, 73, and 112 cm$^{-1}$ (Figure 3d and Figure S13). Experimentally, NCs showed bands at 47, 73, and 126 cm$^{-1}$, in good agreement with theory, though the 65 cm$^{-1}$ mode overlaps with the 73 cm$^{-1}$ feature, appearing as a broadened band at room temperature. This correspondence indicates consistent vibrational behavior between NPLs and the bulk. Analysis of polarizability tensors assigns the 45 cm$^{-1}$ mode to an Ag symmetry mode with Br–Pb–Br scissoring and Pb–Br–Pb bending along z (α angle), the 65 cm$^{-1}$ Ag mode to a related scissoring and bending with distinct distortion, the 73 cm$^{-1}$ B1g mode to Pb–Br–Pb bending within the xy plane, and the 112 cm$^{-1}$ Ag mode to symmetric Pb–Br stretching (Figure 3f).

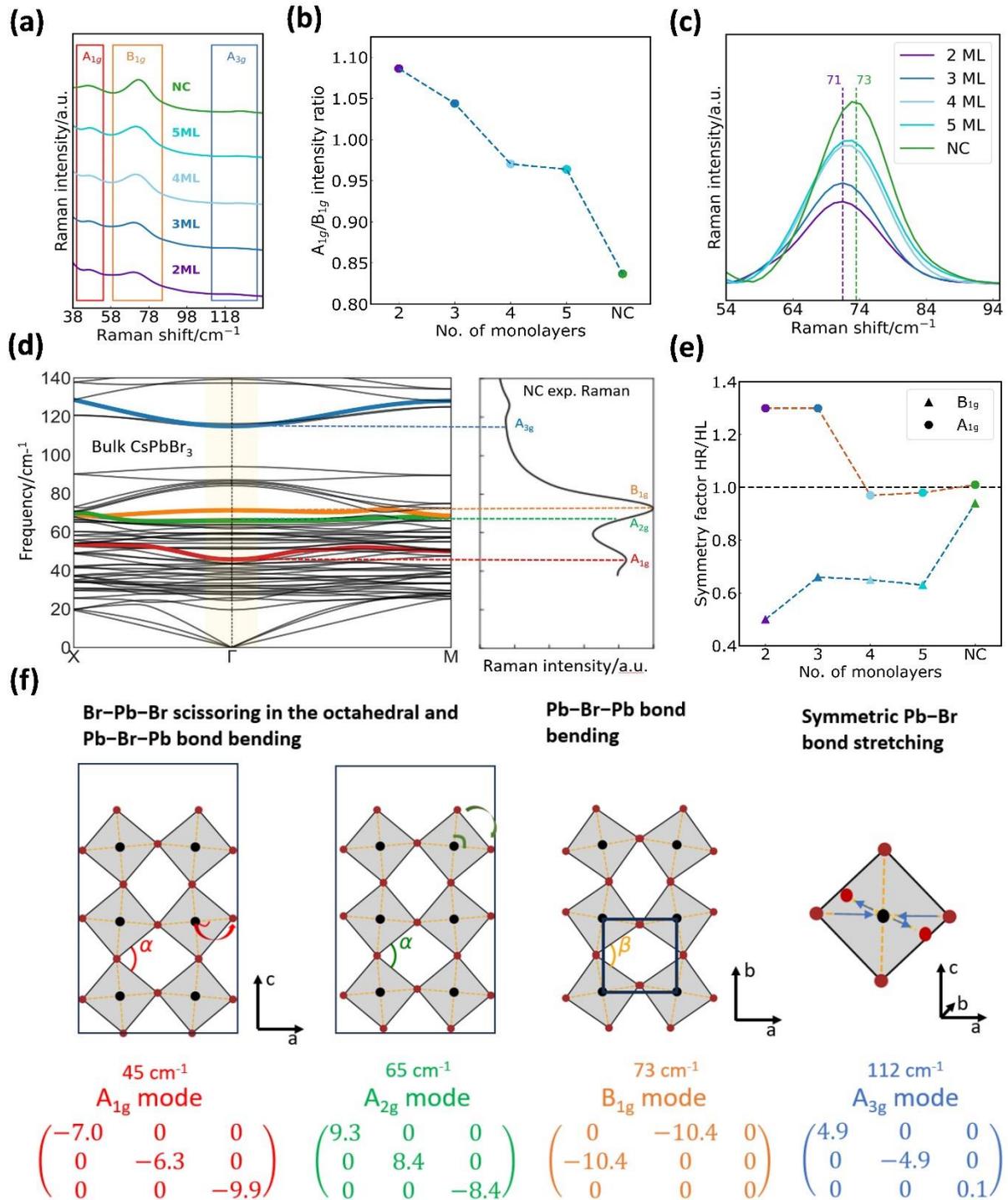

*Figure 3: (a) Raman spectra for the 2ML, 3ML, 4ML, 5ML NPLs, and NCs. Here, emphasis is given to the observed three main peaks at 47 cm$^{-1}$, 73 cm$^{-1}$ and weak feature at around 126 cm$^{-1}$. (b) Mode intensity ratio ($A_{1g}$ / $B_{1g}$) for each NPLs and the NC. (c) Close look on the Raman spectra around mode $B_{1g}$, showing a red shift of the mode frequency with decreased number of MLs. (d) Calculated phonon band structure of orthorhombic CsPbBr$_3$, highlighting the Raman-active modes at the Γ point. (e) Symmetry factor of $A_{1g}$ and $B_{1g}$ modes and (f) Vibrational eigenmodes ($A_{1g}$, $A_{2g}$, $A_{3g}$, and $B_{1g}$) with atomic displacement patterns. The Raman tensor components for each mode are shown, emphasizing their symmetry and selection rules.*

To capture thickness effects, Raman spectra of CsPbBr$_3$ NPLs were modeled using mode-resolved polarizability tensors and slab models (Figure 4a, see SI for formalism). In backscattering geometry, parallel channels (xx, yy) predominantly probe A$_g$ modes, while the crossed channel (xy) isolates B$_{1g}$. As thickness increases from 2 ML to 5 ML, overall intensity grows due to layer accumulation. Experimentally, peaks at 47 cm$^{-1}$ and 71-73 cm$^{-1}$ align well with the theoretical A$_{1g}$ (xx) and B$_{1g}$ (xy) modes at 45 cm$^{-1}$ and 73 cm$^{-1}$ (Figure 3a, 3f), with the strongest enhancement in intensity for B$_{1g}$ mode as thickness grows (Figure 3c).

A quantitative bar plot of the absolute peak-difference metric, $|\Delta I| = |I_{xx}^{Ag} - I_{xy}^{B1g}|$, referenced to the xx low-frequency band, is shown in Figure 4b. Ag modes coupled with xx exhibit no consistent growth, with 5 ML showing even smaller differences than 4 ML, whereas the B$_{1g}$ differences increase steadily. Normalizing each spectrum by layer count (Figure 4b, right) reveals that per-layer growth in xx channels is slightly reduced, while xy continues to grow monotonically from 2 ML to 5 ML, highlighting a distinct symmetry. The weak experimental broad feature at around 125 cm$^{-1}$ for NCs, successively disappearing in the lower dimensional NPLs, aligns with the low intensity A$_{3g}$ mode found at around 120 cm$^{-1}$ in theory, symmetrically expressed only for xx and yy polarization.

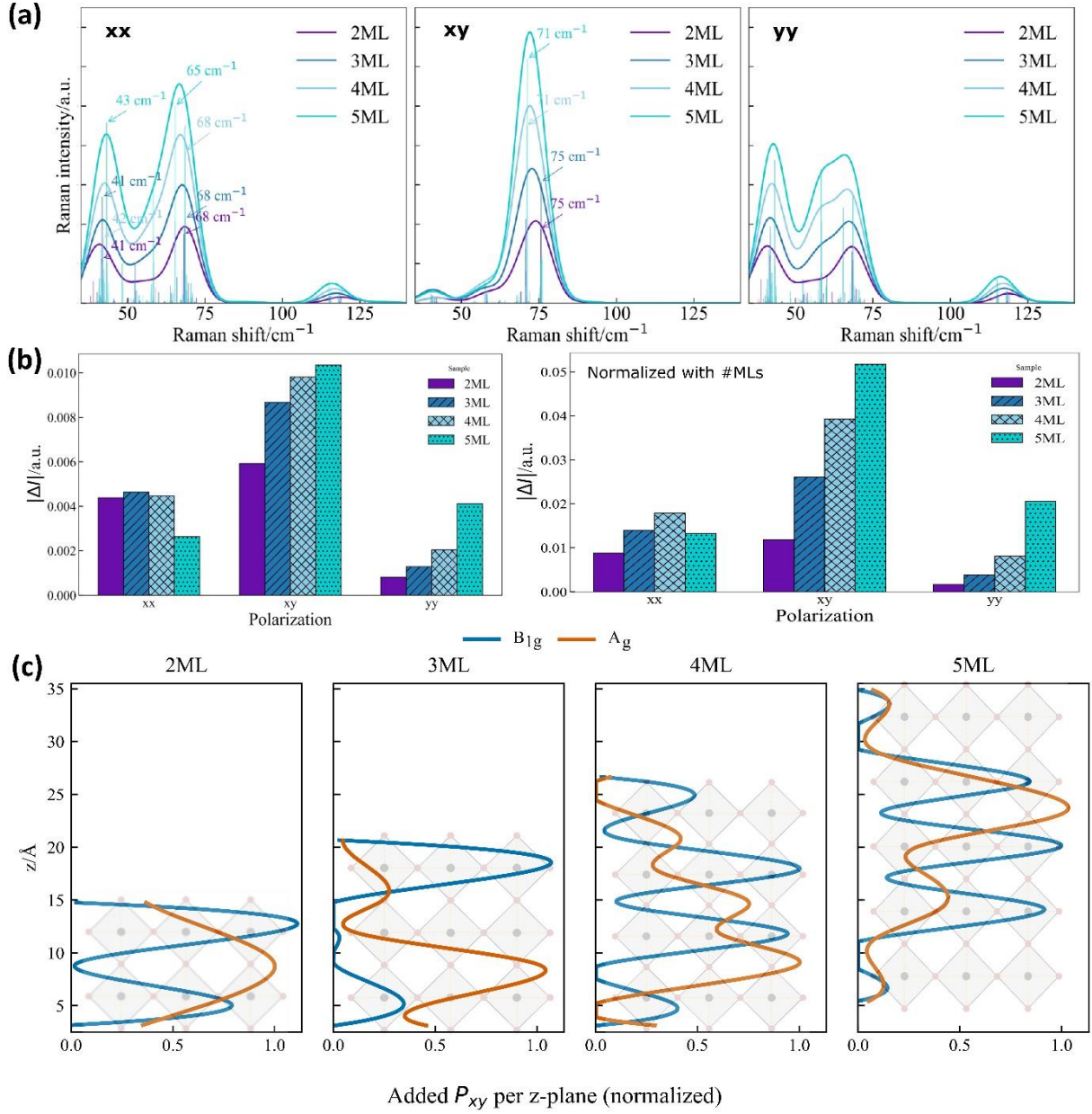

*Figure 4: (a) Polarization-resolved Raman spectra (xx, xy, yy) for 2–5 MLs, highlighting systematic intensity changes of the main phonon bands. (b) Quantitative analysis of thickness-dependent Raman intensity variations shows that $B_{1g}$ modes grow faster with thickness, as calculated (left) and normalized with the number of MLs (right). (c) DFT-calculated plane-resolved polarizability contributions $P_{xy}$ for $B_{1g}$ and $A_g$ modes across thicknesses, revealing that $B_{1g}$ displacements are confined within interior octahedra, while $A_g$ modes involve surface-connected distortions (parameter α), explaining their slower intensity growth with layer number.*

**Discussion**

The intensity ratio between selected Raman modes can be quantitatively explained using the phonon confinement model (PCM), first proposed by Richter et al.[48] and refined by Campbell and Fauchet.[49] PCM describes how reducing crystal dimensions to the nanoscale relaxes the first-order Raman selection rule, which in bulk restricts scattering to zone-center phonons (q = 0). In quantum dots, as shown for ZnO by Raymand et al.,[19] finite lattice size confines the phonon wavefunction, broadening its momentum distribution and enabling contributions from q ≠ 0 states. An equation for the intensity and shape of the phonon modes via the relaxation of the q = 0 condition can be formulated as [19]

$$I(\omega,d) = A \int \frac{\exp\left(-\frac{q^2 d^2}{16\pi^2}\right)}{\left[\omega(0) - \Delta\omega \cdot \sin^2\left(\frac{qa}{4}\right) - \omega(q)\right]^2 + \left(\frac{\Gamma_0}{2}\right)^2} d^3q \quad (1)$$

where $I$ is Raman intensity, $\omega$ vibrational frequency, $d$ particle diameter, $A$ an intensity prefactor depending on the thickness and cross-section of the polarizability of the bonds in the sample, $\omega(0)$ zone center phonon frequency, $\Delta\omega$ the difference between the zone centre and the zone boundary frequency, $\omega(q)$ phonon dispersion, $\Gamma_0$ natural full width at half maximum, $q$ wave vector, and $a$ the lattice parameter in the direction of the vibrational mode. The integration is performed over the whole Brillouin zone. For a 2D system, the phonon dispersion $\omega(q)$ has a functional form of $\omega^2(k_x k_y) = \frac{k_x + k_y}{m}$ leading to a reduced form of the expression with integration over $d^2q$, neglecting $k_z$. In our case, however, as we start from a 2ML system with successively larger extension to 3, 4, and 5 MLs in the $k_z$ direction, the effects from extension in the z-direction are vital to describe the successive change, where a 3D description is required with varied restriction of material dimensions in the z-direction.

Although Equation 1 represents an idealized situation, it provides a starting point to analyse general trends. Since the phonon frequency $\omega(q)$ varies across the Brillouin zone and considering the relaxation of the *q = 0* selection rule for quantum confined materials, one expects a decrease of the phase coherence and a dimensionally induced peak broadening and frequency shifts determined by the dispersion. Negative dispersion ($\Delta\omega < 0$) yields asymmetric, red-shifted tails, while positive dispersion ($\Delta\omega > 0$) gives blue-shifted tails. In bulk $CsPbBr_3$, the $B_{1g}$ branch shows a slight negative dispersion near the Brillouin-zone center (Figure 3d), so PCM predicts a confinement-induced red shift towards lower wavenumbers. Indeed, 4 ML, 5 ML, and NC samples exhibit $B_{1g}$ broadening toward lower wavenumbers for 2 MLs, with a small shift (2 cm$^{-1}$), indicating a small dispersion of the phonon bands as the q=0 selection rule is relaxed. However, in 2 ML and 3 ML samples the $B_{1g}$ mode instead shows asymmetric broadening slightly toward higher wavenumbers (Figure 3c). Our DFT calculations reproduce this anomaly (Figure 4a): in ultrathin slabs the phononic wavevector localizes at the surfaces (Figure S16), shifting the $B_{1g}$ peak upwards. The main $B_{1g}$ peaks appear at 73 cm$^{-1}$ for NC, 71.3 and 71.0 cm$^{-1}$ for 5 ML and 4 ML, but shift to 75.6 cm$^{-1}$ for 3 ML and 2 ML for the theoretical peaks. Thus, while PCM links dispersion sign to asymmetry, surface contributions dominate at low thickness. This is captured experimentally by the symmetry factor S: values >1 reflect higher-wavenumber tails in 2–3 MLs, while values <1 indicate lower-wavenumber broadening in 4–5 MLs and NCs (Figure 3e).

The $A_{1g}$ mode behaves differently. In bulk $CsPbBr_3$ it has positive dispersion (Figure 3d), so PCM predicts a blue shift under confinement. Yet our Raman spectra reveal no frequency shift, only consistent red-tail broadening with decreasing thickness, also reflected in S (Figure 3e). This departure from PCM resembles $TiO_2$ NCs, where the $E_g$ mode follows PCM but the $A_g$ mode does not.[16] Similar failures in rutile $TiO_2$ were attributed to nonstoichiometry and dielectric effects in surface-dominated structures. Such effects are likely here as well, since ultrathin $CsPbBr_3$ NPLs have large surface fractions where disorder and ligand interactions dominate, obscuring the PCM trend for $A_g$.

The distinct behaviors can be rationalized by symmetry. By summing in-plane eigenvector weights $\sum(e_x^2 + e_y^2)$ across z-layers, we mapped each mode's spatial profile (Pxy) (Figure 4c). Across thicknesses, $B_{1g}$ modes concentrate in interior $PbBr_6$ layers within the xy plane, while Ag modes place more weight at xy surfaces, connecting layers via α (the Pb–Br–Pb angle along z). This provides a spatial interpretation: bulk $B_{1g}$ arises from antisymmetric in-plane bending, while Ag mixes in-plane scissoring with out-of-plane coupling. With dimensional reduction, $A_g$ modes relying on z-coupling are disrupted by surfaces, while $B_{1g}$ modes remain protected. Consequently, $B_{1g}$ intensity grows more strongly with thickness, and the cross-polarized xy channel, sensitive to $B_{1g}$, shows the largest thickness dependence. The Pxy metrics directly explain the observed intensity ratio.

A mesoscopic effect further contributes. A 2 ML film behaves as a well-ordered superlattice with clear orientation, favouring one scattering direction. Several works report superlattice formation in ultrathin perovskite NPLs.[4,40,41] This is consistent with multiple GIXRD reflections corresponding to lamellar spacings, characteristic of stacked NPLs.[32] With increasing thickness, twinned or rotated domains and azimuthal disorder enhance xy-active scattering. Thus, intensity evolution arises from intrinsic symmetry changes (interlayer coupling) and extrinsic domain formation that favours cross-polarized response.

Thickness readout in $CsPbBr_3$ nanoplatelets is governed by symmetry-dependent intensity trends, with frequency shifts leaving only a subtle signature. In graphene, the number of layers is determined from the evolution of the 2D band, which splits and changes shape with thickness.[50] In $MoS_2$ and related transition-metal dichalcogenides (TMDCs), the thickness dependence is tracked through the frequency separation between the microscopic in-plane $E_{2g}$ and out-of-plane $A_{1g}$ modes, which increases systematically with interlayer coupling.[51] Both systems rely primarily on frequency shifts as the structural marker of dimensionality. In $CsPbBr_3$ nanoplatelets, however, the Raman bands shift only weakly with thickness, and the decisive fingerprint is the symmetry-selective Raman intensity evolution. Consequently, while TMDCs exploit frequency shifts as thickness probes, halide perovskite nanoplatelets exhibit a distinct symmetry- and localization-driven Raman mechanism to explain the relative intensity ratio with thickness. In graphene, with two atoms in the basis, intensity ratios between the graphite peak (G, $sp^2$) and the first overtone (2D) of the absent diamond peak (D, $sp^3$) are utilized as monolayer metrology for 1-3 MLs[52], while determinations of MLs beyond 3 is challenging. Here, we show that one can go beyond this limit via analysis of the direction dependent polarizabilities in low-dimensional materials to enable ML metrology for a higher number of MLs also in more complex structures as cubic and orthorhombic halide perovskites with 5 atoms or more in the basis.

**Summary**

The crystallographic analysis reveal an energetically allowed coexistence of the cubic and orthorhombic phase for 2 ML $CsPbBr_3$, successively converging to the expected orthorhombic phase of bulk $CsPbBr_3$ upon increased number of MLs. By using polarization-resolved Raman spectroscopy and first-principles calculations, we showed how dimensional confinement reshapes lattice vibrations in $CsPbBr_3$ nanoplatelets. After validating structure and optical confinement, we uncovered a symmetry trend: $B_{1g}$ in-plane modes intensify and broaden with red tails, consistent with a phonon confinement model, whereas $A_{1g}$ out-of-plane modes remain less affected. DFT explains this by interior localization for the former and z-dependent features for the latter, especially in 2–3 ML platelets where surface phonons dominate. Practically, this symmetry-resolved Raman fingerprint provides a calibrated, non-destructive thickness metrology utilizing the $A_{1g}/B_{1g}$ intensity ratio for determination of the number of monolayers. Interior-localized vibrations couple differently to carriers and heat than surface vibrations, with implications for polaron formation, hot-phonon bottlenecks, thermal transport, and stability. The symmetry framework also provides design knobs: thickness, dielectric environment, surface termination, strain, and composition can be tuned to steer localization and tailor exciton–phonon interactions. These insights are directly relevant to optoelectronic performance and reliability.

The analysis is based on idealized slabs and harmonic vibrations, while interfacial roughness, defects, or anharmonicity can be added without loss of generality. Altogether, we provide a mechanistic framework and practical toolkit to understand lattice dynamics and corresponding phonons in low-dimensional perovskites. While demonstrated for 2D $CsPbBr_3$ perovskites with successive growth of the number of MLs, the symmetry-driven trends can be anticipated to hold also for other 2D materials as long as there are one in-plane and one out-of-plane phonon mode detectable in Raman, IR, Inelastic Neutron Scattering, or (Resonant) Inelastic X-ray Scattering experiments. More broadly, the principles outlined here should be transferable to a wider class of quantum-confined lattices and vibration spectroscopies where symmetry and dimensionality dictate vibrational properties.

**Materials, Methods, Characterization, and Computational Details**

The preparation of $CsPbBr_3$ nanoplatelets followed the approach by Bohn et. al. [47] with some modifications.

### Materials

$Cs_2CO_3$ (ceasium carbonate, 99\%), $PbBr_2$ (lead (II) bromide, 98\%), oleic acid (technical grade 90\%), oleylamine (technical grade 70\%), toluene (for HPLC, 99.9\%), acetone (for HPLC, 99.9\%), hexane (for HPLC, 97.0\%, GC) and were purchased from Sigma-Aldrich.

### Preparation of precursors

Prior to precursor preparation, both oleic acid and oleylamine were pre-heated to 110 °C and subsequently filtered through a 0.2 μm syringe filter to remove any precipitated impurities. The cesium oleate (Cs-oleate) precursor was synthesized by dissolving 0.1 mmol of $Cs_2CO_3$ powder in 10 mL of oleic acid at 100 °C under continuous stirring until complete dissolution. The $PbBr_2$ precursor solution was prepared separately by dissolving 0.1 mmol of $PbBr_2$ along with 100 μL each of oleylamine and oleic acid in 10 mL of toluene at 100 °C. The coordinating ligands (oleic acid and oleylamine) facilitate the dissolution of $PbBr_2$ in the non-polar toluene medium. Both precursor solutions were subsequently transferred to and stored in a nitrogen-filled glovebox for further use.

### Synthesis of CsPbBr3 NPLs

All synthesis steps were conducted at room temperature in a standard fume hood. Control over the nanoplatelet (NPL) thickness was achieved by adjusting the volume ratios of the two precursors and the antisolvent employed during precipitation. In a typical procedure, varying amounts of the Cs-oleate precursor were rapidly injected into the $PbBr_2$–ligand precursor solution under vigorous stirring to ensure homogenous nucleation and complete crystallization of the NPLs. After 5 seconds, acetone was added in specific volumes as an antisolvent to promote NPL precipitation from the toluene phase. Following 1 minute of stirring, the mixture was centrifuged at 4000 rpm for 3 minutes, and the resulting precipitate was redispersed in hexane to yield a stock solution with a concentration of 15 mg/mL.

At this stage, the solution was suitable for characterization; however, to improve colloidal stability and eliminate larger crystalline aggregates, a second centrifugation step was performed at 3000 rpm for 3 minutes without adding antisolvent. The supernatant was then carefully extracted using a syringe to minimize disturbance. Accurate determination of the final concentration was achieved through multiple weight measurements of the centrifuge tubes. The resulting stock solution was subsequently diluted to a working concentration of 5 mg/mL, which was used throughout the remainder of the study unless stated otherwise.

The optimized precursor volume ratios for obtaining NPLs with 2, 3, 4, and 5 monolayers (MLs) were as follows (Cs-oleate/PbBr$_2$–ligand/acetone in μL/mL/mL): 150/3/2 for 2 MLs, 150/1.5/2 for 3 MLs, 150/1.2/2 for 4 MLs, and 200/1/2 for 5 MLs. For the 5 ML NPLs, an additional 0.2 mL of acetone was pre-mixed with the PbBr$_2$–ligand precursor prior to Cs-oleate injection to suppress the formation of bulk crystals during nucleation.

### Characterization

All characterization measurements, unless otherwise specified, were performed on drop-cast films prepared from NPL dispersions in hexane, deposited onto glass or gold substrates. Photoluminescence (PL) enhancement and PL quantum yield (PLQY) measurements were conducted in solution at matched concentrations to ensure consistency.

PL spectra were acquired using a Renishaw inVia confocal PL–Raman system equipped with a 405 nm excitation laser (FWHM ≈ 1 nm). UV–Vis absorption spectra were collected using an Agilent Cary 5000 spectrophotometer equipped with halogen and deuterium lamps, from films deposited on glass substrates. Raman spectroscopy was carried out on drop-cast films on gold-coated glass substrates using the Renishaw inVia confocal system with a 785 nm excitation source. A 50× objective lens was employed, and the laser power was set to 83 mW, corresponding to a power density of approximately $5.3 \times 10^7$ mW/mm², based on a 1.4 μm laser spot size. This power density was the maximum threshold that could be applied without inducing degradation of the perovskite NPLs. Raman spectra were collected in both extended and static modes with an acquisition time of 10 s per scan, and a total of three accumulations per sample.

Scanning Transmission electron microscopy (STEM) was performed on fresh NPLs samples prepared by drop-casting hexane-dispersed NPLs (1 mg/mL) onto holey-carbon Cu grids (Ted Pella, 200 mesh). Samples were prepared within a few hours before analysis, air-dried at room temperature. STEM imaging was conducted using a Cs-probe corrected FEI Titan Themis 200 microscope equipped with a Schottky field emission gun and operated at an accelerating voltage of 200 kV. Beam current was maintained between 20 and 70 pA in STEM mode.

Grazing incidence X-ray diffraction (GIXRD) measurements were carried out on NPL films drop-cast on SiO$_x$/Si substrates using a Malvern PANalytical Empyrean system equipped with a copper X-ray source. The incidence angle was fixed at 0.5° for all samples to minimize background scattering from the substrate.

### Density Functional Theory Calculations

Density Functional Theory (DFT) calculations were carried out utilizing the Projected Augmented Wave (PAW) approach within the Vienna Ab initio Simulation Package (VASP) [53,54]. The generalized gradient approximation with the Perdew, Burke, and Ernzerhof (PBE) parametrization to address the exchange and correlation terms within the Kohn-Sham Hamiltonian [55] was employed (unless specified) with the addition of spin-orbit coupling (SOC). Plane waves were expanded to a cutoff of 400 eV (unless specified), and the Brillouin zone was sampled with different grids depending on the structural lattice (3x3x1 for the orthorhombic and 2x2x1 for cubic distorted slabs and tetragonal).

### Vibrational Modes and Raman Activities

To examine the phonon modes and frequencies at the Γ point, density functional perturbation theory (DFPT) was employed. During this phase, the employed structure underwent optimization with a force convergence target of 0.001 eV/Å and an energy convergence goal of $10^{-8}$ eV. Phonon band structure and density of states were computed using Phononpy [56].

Raman spectra were calculated using polarizability derivatives within the non-resonant Placzek approximation—a well-established approach in ab initio Raman modeling [57,58]. Mode amplitudes were obtained by projecting the Raman tensor onto the incident and detected polarization vectors, with intensities weighted by thermal occupation and phonon frequency at 300 K. Polarization configurations (xx, xy, yy) were defined to isolate symmetry-allowed scattering channels. The resulting discrete lines were broadened using Gaussian functions to mimic instrumental linewidths, producing continuous spectra suitable for comparison with experiment. Full methodological details and convergence tests are given in the Supplementary Information.

**Supporting Information**

Supplementary information is available in the online version, including additional details on the theoretical background, structures used in the DFT calculations, polarizability tensors for the Raman calculations, the Raman and PL of the post-treatment approach, the Raman spectra of the Iodide system, the TEM images and electron diffraction analysis, and the methods and codes for the calculation of the symmetrical parameter used in X-ray diffraction and Raman sections.

**Code Availability**

The VASP code is licensed software available from https://www.vasp.at/.

**Acknowledgments**

We thank the Swedish Energy Agency (P2020-90215), FORMAS (Grant no 2022-02297), and the Swedish Research Council (Grant no 2023-05244) for financial support. We acknowledge Myfab Uppsala for providing facilities and experimental support. Myfab is funded by the Swedish Research Council (2020-00207) as a national research infrastructure. The computations were performed using resources provided by the National Academic Infrastructure for Supercomputing in Sweden (NAISS) through project NAISS 2024/5-372 and 2025/5-472.

**Reference**

1. Weidman, M. C., Seitz, M., Stranks, S. D. & Tisdale, W. A. Highly Tunable Colloidal Perovskite Nanoplatelets through Variable Cation, Metal, and Halide Composition. *ACS Nano* **10**, 7830–7839 (2016).

2. Edvinsson, T. Optical quantum confinement and photocatalytic properties in two-, one- and zero-dimensional nanostructures. *R. Soc. open sci.* **5**, 180387 (2018).

3. Sapori, D., Kepenekian, M., Pedesseau, L., Katan, C. & Even, J. Quantum confinement and dielectric profiles of colloidal nanoplatelets of halide inorganic and hybrid organic–inorganic perovskites. *Nanoscale* **8**, 6369–6378 (2016).

4. Movilla, J. L., Planelles, J. & Climente, J. I. Excitons in metal halide perovskite nanoplatelets: an effective mass description of polaronic, dielectric and quantum confinement effects. *Nanoscale Adv.* **5**, 6093–6101 (2023).

5. Wang, Q. *et al.* Quantum confinement effect and exciton binding energy of layered perovskite nanoplatelets. *AIP Advances* **8**, 025108 (2018).

6. Gramlich, M., Lampe, C., Drewniok, J. & Urban, A. S. How Exciton–Phonon Coupling Impacts Photoluminescence in Halide Perovskite Nanoplatelets. *J. Phys. Chem. Lett.* **12**, 11371–11377 (2021).

7. Wright, A. D. *et al.* Electron–phonon coupling in hybrid lead halide perovskites. *Nat Commun* **7**, 11755 (2016).

8. Strong Carrier–Phonon Coupling in Lead Halide Perovskite Nanocrystals | ACS Nano. https://pubs.acs.org/doi/10.1021/acsnano.7b05033.

9. Akkerman, Q. A. *et al.* Solution Synthesis Approach to Colloidal Cesium Lead Halide Perovskite Nanoplatelets with Monolayer-Level Thickness Control. *J. Am. Chem. Soc.* **138**, 1010–1016 (2016).

10. Harnessing Hot Phonon Bottleneck in Metal Halide Perovskite Nanocrystals via Interfacial Electron–Phonon Coupling | Nano Letters. https://pubs.acs.org/doi/full/10.1021/acs.nanolett.0c01452.

11. Ghosh, T., Aharon, S., Etgar, L. & Ruhman, S. Free Carrier Emergence and Onset of Electron–Phonon Coupling in Methylammonium Lead Halide Perovskite Films. *J. Am. Chem. Soc.* **139**, 18262–18270 (2017).

12. Straus, D. B. *et al.* Direct Observation of Electron–Phonon Coupling and Slow Vibrational Relaxation in Organic–Inorganic Hybrid Perovskites. *J. Am. Chem. Soc.* **138**, 13798–13801 (2016).

13. Guo, Z., Wu, X., Zhu, T., Zhu, X. & Huang, L. Electron–Phonon Scattering in Atomically Thin 2D Perovskites. *ACS Nano* **10**, 9992–9998 (2016).

14. Zhao, D. *et al.* Monitoring Electron–Phonon Interactions in Lead Halide Perovskites Using Time-Resolved THz Spectroscopy. *ACS Nano* **13**, 8826–8835 (2019).

15. Madito, M. J. Revisiting the Raman disorder band in graphene-based materials: A critical review. *Vibrational Spectroscopy* **139**, 103814 (2025).

16. Mazza, T. *et al.* Raman spectroscopy characterization of $\mathrm{Ti}{\mathrm{O}}_{2}$ rutile nanocrystals. *Phys. Rev. B* **75**, 045416 (2007).

17. Swamy, V. Size-dependent modifications of the first-order Raman spectra of nanostructured rutile ${\text{TiO}}_{2}$. *Phys. Rev. B* **77**, 195414 (2008).

18. Swamy, V., Muddle, B. C. & Dai, Q. Size-dependent modifications of the Raman spectrum of rutile TiO2. *Appl. Phys. Lett.* **89**, 163118 (2006).

19. Raymand, D., Jacobsson, T. J., Hermansson, K. & Edvinsson, T. Investigation of Vibrational Modes and Phonon Density of States in ZnO Quantum Dots. *J. Phys. Chem. C* **116**, 6893–6901 (2012).

20. Cançado, L. G. *et al.* Anisotropy of the Raman Spectra of Nanographite Ribbons. *Phys. Rev. Lett.* **93**, 047403 (2004).

21. Zhang, K. *et al.* Raman signatures of inversion symmetry breaking and structural phase transition in type-II Weyl semimetal MoTe2. *Nat Commun* **7**, 13552 (2016).

22. Dhakal, K. P. *et al.* Giant Modulation of Interlayer Coupling in Twisted Bilayer ReS2. *Advanced Science* **12**, 2500411 (2025).

23. Kirschner, M. S. *et al.* Photoinduced, reversible phase transitions in all-inorganic perovskite nanocrystals. *Nat Commun* **10**, 504 (2019).

24. Yaffe, O. *et al.* Local Polar Fluctuations in Lead Halide Perovskite Crystals. *Phys. Rev. Lett.* **118**, 136001 (2017).

25. Mannino, G. *et al.* Temperature-Dependent Optical Band Gap in CsPbBr3, MAPbBr3, and FAPbBr3 Single Crystals. *J. Phys. Chem. Lett.* **11**, 2490–2496 (2020).

26. Wen, J.-R., Rodríguez Ortiz, F. A., Champ, A. & Sheldon, M. T. Kinetic Control for Continuously Tunable Lattice Parameters, Size, and Composition during CsPbX3 (X = Cl, Br, I) Nanorod Synthesis. *ACS Nano* **16**, 8318–8328 (2022).

27. dos Reis, R. *et al.* Determination of the structural phase and octahedral rotation angle in halide perovskites. *Appl. Phys. Lett.* **112**, 071901 (2018).

28. Whitcher, T. J. *et al.* Dual phases of crystalline and electronic structures in the nanocrystalline perovskite CsPbBr3. *NPG Asia Mater* **11**, 70 (2019).

29. Cottingham, P. & Brutchey, R. L. On the crystal structure of colloidally prepared CsPbBr3 quantum dots. *Chem. Commun.* **52**, 5246–5249 (2016).

30. Brennan, M. C., Kuno, M. & Rouvimov, S. Crystal Structure of Individual CsPbBr3 Perovskite Nanocubes. *Inorg. Chem.* **58**, 1555–1560 (2019).

31. Hooper, T. J. N., Fang, Y., Brown, A. A. M., Pu, S. H. & White, T. J. Structure and surface properties of size-tuneable CsPbBr3 nanocrystals. *Nanoscale* **13**, 15770–15780 (2021).

32. Diroll, B. T., Banerjee, P. & Shevchenko, E. V. Optical anisotropy of CsPbBr3 perovskite nanoplatelets. *Nano Convergence* **10**, 18 (2023).

33. Bertolotti, F. *et al.* Coherent Nanotwins and Dynamic Disorder in Cesium Lead Halide Perovskite Nanocrystals. *ACS Nano* **11**, 3819–3831 (2017).

34. Bertolotti, F. *et al.* Crystal Structure, Morphology, and Surface Termination of Cyan-Emissive, Six-Monolayers-Thick CsPbBr3 Nanoplatelets from X-ray Total Scattering. *ACS Nano* **13**, 14294–14307 (2019).

35. Shamsi, J. *et al.* Stable Hexylphosphonate-Capped Blue-Emitting Quantum-Confined CsPbBr3 Nanoplatelets. *ACS Energy Lett.* **5**, 1900–1907 (2020).

36. Udayabhaskararao, T. *et al.* A Mechanistic Study of Phase Transformation in Perovskite Nanocrystals Driven by Ligand Passivation. *Chem. Mater.* **30**, 84–93 (2018).

37. Toso, S. *et al.* Multilayer Diffraction Reveals That Colloidal Superlattices Approach the Structural Perfection of Single Crystals. *ACS Nano* **15**, 6243–6256 (2021).

38. Toso, S., Baranov, D., Filippi, U., Giannini, C. & Manna, L. Collective Diffraction Effects in Perovskite Nanocrystal Superlattices. *Acc. Chem. Res.* **56**, 66–76 (2023).

39. Zhang, H., Chen, B. & F. Banfield, J. The size dependence of the surface free energy of titania nanocrystals. *Physical Chemistry Chemical Physics* **11**, 2553–2558 (2009).

40. Müller, E., Vogelsberger, W. & Fritsche, H.-G. The dependence of the surface energy of regular clusters and small crystallites on the particle size. *Crystal Research and Technology* **23**, 1153–1159 (1988).

41. Samsonov, V. M., Sdobnyakov, N. Y. & Bazulev, A. N. Size dependence of the surface tension and the problem of Gibbs thermodynamics extension to nanosystems. *Colloids and Surfaces A: Physicochemical and Engineering Aspects* **239**, 113–117 (2004).

42. Fritsche, H.-G. Particle Size Effects in Pair Potential Approximation II. Particles of Ion Pairs. *Zeitschrift für Physikalische Chemie* **189**, 53–62 (1995).

43. Wang, L. *et al.* Ultra-stable CsPbBr3 Perovskite Nanosheets for X-Ray Imaging Screen. *Nano-Micro Lett.* **11**, 52 (2019).

44. Dang, Z. *et al.* In Situ Transmission Electron Microscopy Study of Electron Beam-Induced Transformations in Colloidal Cesium Lead Halide Perovskite Nanocrystals. *ACS Nano* **11**, 2124–2132 (2017).

45. Chen, S. & Gao, P. Challenges, myths, and opportunities of electron microscopy on halide perovskites. *J. Appl. Phys.* **128**, 010901 (2020).

46. Protesescu, L. *et al.* Nanocrystals of Cesium Lead Halide Perovskites (CsPbX3, X = Cl, Br, and I): Novel Optoelectronic Materials Showing Bright Emission with Wide Color Gamut. *Nano Lett.* **15**, 3692–3696 (2015).

47. Bohn, B. J. *et al.* Boosting Tunable Blue Luminescence of Halide Perovskite Nanoplatelets through Postsynthetic Surface Trap Repair. *Nano Lett.* **18**, 5231–5238 (2018).

48. Richter, H., Wang, Z. P. & Ley, L. The one phonon Raman spectrum in microcrystalline silicon. *Solid State Communications* **39**, 625–629 (1981).

49. Campbell, I. H. & Fauchet, P. M. The effects of microcrystal size and shape on the one phonon Raman spectra of crystalline semiconductors. *Solid State Communications* **58**, 739–741 (1986).

50. Ferrari, A. C. & Basko, D. M. Raman spectroscopy as a versatile tool for studying the properties of graphene. *Nature Nanotech* **8**, 235–246 (2013).

51. Lee, C. *et al.* Anomalous Lattice Vibrations of Single- and Few-Layer MoS2. *ACS Nano* **4**, 2695–2700 (2010).

52. Raman Spectroscopy: From Graphite to sp2 Nanocarbons. in *Raman Spectroscopy in Graphene Related Systems* 73–101 (John Wiley & Sons, Ltd, 2011). doi:10.1002/9783527632695.ch4.

53. Kresse, G. & Furthmüller, J. Efficient iterative schemes for ab initio total-energy calculations using a plane-wave basis set. *Phys. Rev. B* **54**, 11169–11186 (1996).

54. Kresse, G. & Joubert, D. From ultrasoft pseudopotentials to the projector augmented-wave method. *Phys. Rev. B* **59**, 1758–1775 (1999).

55. Perdew, J. P., Burke, K. & Ernzerhof, M. Generalized Gradient Approximation Made Simple. *Phys. Rev. Lett.* **77**, 3865–3868 (1996).

56. Togo, A., Chaput, L., Tadano, T. & Tanaka, I. Implementation strategies in phonopy and phono3py. *J. Phys.: Condens. Matter* **35**, 353001 (2023).

57. Walter, M. & Moseler, M. Ab Initio Wavelength-Dependent Raman Spectra: Placzek Approximation and Beyond. *J. Chem. Theory Comput.* **16**, 576–586 (2020).

58. Khoo, K. H. & Chelikowsky, J. R. First-principles study of vibrational modes and Raman spectra in P-doped Si nanocrystals. *Phys. Rev. B* **89**, 195309 (2014).